\documentclass[12pt]{article}
\usepackage{amsfonts,amsbsy}
\newcommand{\Epsilon}{{\cal E}}
\newcommand{\be}{\begin{equation}}
\newcommand{\ee}{\end{equation}}

\newcommand{\SUN}{SU(N)}

\newcommand{\SUT}{SU(2)}

\newcommand{\bea}{\begin{eqnarray}}
\newcommand{\eea}{\end{eqnarray}}
\input{epsf.tex}

\begin{document}

\vskip -4cm

\begin{flushright}
FTUAM-98-17

IFT-UAM/CSIC-98-12
\end{flushright}

\vskip 0.2cm

{\Large
\centerline{{\bf Self-dual vortex-like  configurations}}
\centerline{{\bf in SU(2) Yang-Mills Theory}}
\vskip 0.3cm

\centerline{\qquad \ \ 
  A. Gonz\'alez-Arroyo{$^\dag$}{$^\ddag$} and A. Montero{$^\dag$} \\ }}
\vskip 0.3cm

\centerline{{$^\dag$}Departamento de F\'{\i}sica Te\'orica C-XI,}
\centerline{Universidad Aut\'onoma de Madrid,}
\centerline{Cantoblanco, Madrid 28049, SPAIN.}
\vskip 10pt
\centerline{{$^\ddag$}Instituto de F\'{\i}sica Te\'orica C-XVI,}
\centerline{Universidad Aut\'onoma de Madrid,}
\centerline{Cantoblanco, Madrid 28049, SPAIN.}

\vskip 0.8cm

\begin{center}
{\bf ABSTRACT}
\end{center}
We show that there are  solutions of the SU(2) Yang-Mills classical
equations of motion in $R^4$,  which are self-dual and vortex-like(fluxons).
The action density is concentrated along a thick two-dimensional wall
(the world sheet of a straight infinite vortex line). The configurations
are constructed   from self-dual $R^2 \times T^2$ configurations.

\vskip 1.5 cm
\begin{flushleft}

\end{flushleft}

\newpage

\section{Introduction}
The purpose of this paper is to show that there exist classical self-dual
SU(2) Yang-Mills configurations in $R^4$  which are localized around  a given
2-dimensional sheet. This sheet is the world sheet of
a spatial  flux-line (vortex or fluxon). Being self-dual the configuration is
classically stable. The construction of the configurations proceeds by
naturally embedding (by considering an infinite number of periods) a
field configuration defined in $R^2 \times T^2$ ($T^n$ is the n-dimensional torus)
into one in  $R^4$.

 Some years
ago~\cite{gpg-as},
our group  studied the existence and properties of self-dual Yang-Mills
configurations on the torus by numerical methods. Motivated by the
Hamiltonian formulation, we considered the limit in which one of the
sizes(time) of the torus goes to infinity,
thus producing a well-defined $R \times
T^3$ gauge field configuration which is localized in time. The configuration
that we constructed was self-dual and carried topological charge $\frac{1}{2}$.
Later on, configurations living on $R \times T^3$ with unit topological charge
were  found and studied~\cite{overimp}. 

There are two recent sets 
of works by other authors
which have convinced us of the interest of looking into configurations which
are living in  $R^2 \times T^2$ and localized in the 2 large directions.
The first is a series of papers~\cite{vanbaalnew} in which the authors
were able to give an explicit analytic expression for a whole set of
new caloron solution with non-trivial holonomy. These are self-dual Yang-Mills
configurations in $R^3 \times T^1$ which are localized in the 3 non-periodic
directions. Actually, the action density peaks at two points in $R^3$.
If we naturally  embed the configuration in $R^4$  the action density is
concentrated around 2 lines, which are the world-lines of the so-called
{\em constituent monopoles}. Thus, the case studied here ($R^2 \times T^2$)
is  missing to complete the picture. Of course, our numerical methods are a
poor substitute of the analytic formulas. However, they can provide hints
which could eventually lead to an analytic  expression for these gauge field
configurations. Furthermore, in the absence of analytic expressions our 
numerical method allows the determination, with high precision,  of the
different physical quantities and properties of these configurations. 

A second group of papers~\cite{greensite} which motivated this work,
consists on recent lattice gauge theory studies that give evidence in favour
of the relevance of vortex-like configurations to explain the Confinement
property of Yang-Mills theory. The presence of vortices in typical (generated
by Monte Carlo simulations) Yang-Mills vacuum configurations  is shown  by
performing the maximal center projection. The resulting $Z_2$ 
degrees of freedom can account for the observed string tension, a
behaviour described as {\em center dominance}. The authors (see Ref.~\cite{lat98})
argue that the location of the resulting $Z_2$ vortices
signal the presence of underlying {\em center vortices}. These correspond to
certain action density structures, which
their center projection has allowed to pin down even without {\em
cooling}~\cite{cooling}. In
comparing with the more standard  abelian projection, they argue that the
abelian monopoles are located within the vortex walls, but their exact
location has no special significance: they simply mimic the structure
provided by the vortex wall. The crucial issue is then, what is the
dimensionality of the relevant structures which are responsible for
Confinement. The supporters  of the fluxonic(vortex) scenario would say 2,
as opposed to the unit dimensionality of monopole world-lines. 

How do our
configurations enter the picture? They provide candidates for the underlying
action density structures: {\em center vortices}. As we will see, they indeed 
carry one unit of magnetic center flux~\cite{thooft}. 
Furthermore, the configurations, being self-dual, are solutions of the Euclidean equations
of motion corresponding to a local minimum of the action. For example, they would  
emerge as structures seen by {\em cooling}. Even if this is not the
case, the realization that there are  vortex-like solutions of the classical 
Yang-Mills equations of motion is an important issue in itself.

The plan of the paper is as follows. In the next section we briefly review
our method and present our results. Finally, in the last section, we discuss 
in further detail  a few points concerning the way in which the configurations
could appear in the Yang-Mills vacuum and the possible relation to Confinement.

\section{Numerical method and results}
Our starting point is the consideration of $\SUT$ Yang-Mills fields
(we take $\SUT$ for simplicity, although most statements remain valid,
with the
appropriate changes, for $\SUN$)    on the
4-dimensional torus (for a review see~\cite{review}). `t Hooft~\cite{thooft}
realized
that when putting this fields on the torus, there are some new topological
sectors, known as twist sectors, characterized by 2 integer mod 2 
vectors: $(\vec{m})_i=\frac{1}{2} \epsilon_{i j k} n_{j k}$ and
 $(\vec{k})_i=n_{0 i}$. Thus, gauge fields (actually the bundles) can be
classified
 into different sets labeled by these vectors and the ordinary topological
charge. One can minimize the action functional within each set to obtain
a classical solution (of the Euclidean equations of motion). The absolute
minimum within each sector is bounded by $8 \pi^2$ times the absolute value
of the topological charge, and only saturated by self-dual or anti-self-dual
configurations. Our strategy  to find the solutions is to consider the
lattice formulation of Yang-Mills theory, and perform a minimization of
the lattice action functional (Wilson action in our case). Twist sectors are
easily implemented on the lattice~\cite{gjk-a}. Unfortunately, the set of
lattice fields within each sector ($SU(2)^V$) is a connected set, and there
is no natural splitting into topological charge sectors. Thus, what can be
easily done with our methods is to find the configuration which minimizes the
action for each twist sector without specifying the value of the topological charge.
Particularly interesting are thus the so-called non-orthogonal twists
($\vec{k}\cdot \vec{m} \ne 0 \bmod 2 $) for which the topological charge is half
integer. The resulting configuration is assured to be non-trivial ($F_{\mu
\nu} \ne 0$). For the minimization procedure we use the naive cooling
algorithm~\cite{cooling}. Since we are looking for the absolute minimum
action configuration we do not  have to be too sophisticated in this point.
We refer the interested reader to Ref.~\cite{gpg-as} for further details on the
numerical technique. It is worthwhile to mention, that it  takes only a
few minutes in 
a standard workstation to generate one of the configurations used in this
paper.

The specific characterization of this work compared to
previous ones, is that  we make the torus asymmetric  in the different
directions: 2 of the periods are taken much larger than the  other two. Thus
our lattices are of size $N_l \times N_s \times N_l \times N_s$ with the
order of directions labeled $0,1,2,3$ and $N_l \gg N_s$. If a limiting
 $R^2\times T^2$ configuration exists, we  expect that our lattice
configurations would  tend to a well-defined limit as
$N_l/N_s \rightarrow \infty$. Once this limit is approximately obtained, one
must study the approach to the continuum limit as $a \equiv 1/N_s$ tends to
zero. In this work, we have used   $N_s=4,6$ and values of  $N_l$ ranging
from  $12$ to $24$.

Concerning the value of the twist vectors $\vec{k}$ and $\vec{m}$, all
non-orthogonal values ($\vec{k} \cdot \vec{m} \ne 0 $) have been studied. 
This implies that our resulting configurations have topological charge
$|Q|=\frac{1}{2}$, and
the corresponding action is bounded from below by $4 \pi^2$. Using the
symmetries of the problem, one can reduce to $14$ the total number of twist vector
sets to study.

Now we will present our results. The first quantity which one obtains after
the minimization process, is the value of the  minimum lattice action
$S_L$ for each twist and lattice size. By looking at this quantity alone one
observes that the different sets of twist vectors can be classified into
three classes. Class A is characterized by $m_2=n_{1 3}=1 \bmod 2$.
$S_L$ is very similar for all twist-vectors within this class. We get
$S_L/(4 \pi^2)$ is 0.9597(1) for $N_s=4$ $N_l=16,20$ and 0.982737(1) for
$N_s=6$ and $N_l=24$.  The number in parenthesis affects the last digit and
represents the maximum variation obtained within the class. The fact that
$S_L$ does not depend on $N_l$ to the $10^{-4}$ level (actually less than this,
 by comparing equal twist vectors) is in agreement with our expectation
for a lattice configuration approximating
 an $R^2 \times T^2$ one. The fact that all components of
the twist vectors, except for $m_2=n_{13}$, become  irrelevant in this limit,
is also natural,
 since they involve one or both of the directions which tend to
$R$. Now, we examine the approach of $S_L$ to the continuum limit.
By using our  $N_s=4$ and $N_s=6$ values in the
formula  $S_L=S_c + B a^2$, we determine the 2 parameters to be $B=-0.6634\,
(4 \pi^2)$
and $S_c=1.0011\, (4 \pi^2)$. Thus, the latter value equals  the
continuum action of a (anti-)self-dual configuration to  1 part in $10^3$. This
discrepancy is of the size to be accounted for by  $O(a^4)$ corrections. In
summary, class A  configurations behave according to our expectations,
and provide a natural candidate for our $R^2 \times T^2$ configurations. The
rest of our study will concentrate on them.

 For completeness,  we mention
that the remaining choices of twist vectors (having $m_2=0 \bmod 0$) can be
classified into 2 sets.  Class B is made of the configurations with $n_{0 1}
=n_{0 3}=n_{2 3}$ odd  and $n_{1 2}$ even and all obtained by exchanging $0
\leftrightarrow 2$ and $1 \leftrightarrow 3$. The rest is labeled class C.
Again $S_L$ changes very little within each class. This time, however,
the variation of $S_L$ with $N_l$ is comparable to that with $N_s$.
 This  suggests that the configuration does not
 become localized in $t$ and $y$ as the corresponding periods go to infinity.

Now, we study the action density distribution for all class A configurations.
The different twist vectors  in the class give results which are consistent
modulo space-time translations. The distribution is symmetric under the
exchange $t \leftrightarrow y$ and $x \leftrightarrow z$. Another feature is
that the action density distribution has a unique local maximum.   Indeed,
we use several  interpolation methods to determine the location of the
maximum with a precision of less than a tenth of the lattice spacing in each
direction. Then, we make use of translation invariance to choose the origin
of coordinates precisely at this point. With this choice, the action density
distribution is even under a change of sign  of any coordinate ($x_{\mu}
\rightarrow -x_{\mu}$). In all subsequent expressions we write the value of
the coordinates  on the continuum taking $a=\frac{1}{N_s}$ (The periods in x
and z are set to 1).  

To give an idea of the behavior of the action density distribution,  we studied  the
energy profiles $\Epsilon_{\mu}(x_{\mu})$, which are the integrals of the
action density over 3 of the 4 variables.  $\Epsilon_1=\Epsilon_3$ is a
periodic function of its argument with period $l_1=l_3=1$. If we determine
the first few Fourier coefficients and compare the data to the curve $S_L +4
\pi^2\, B
\cos(2 \pi x) +  4 \pi^2 C\, \cos(4 \pi x)$, we obtain $B=0.1612(3), C=0.004(4)$ for $N_s=4$
and  $B=0.16380(5), C=0.00646(1)$ for $N_s=6$. Errors represent  again the
variation within class A. Indeed, the first 2 Fourier coefficients describe
the data to the $1 \%$ level. A quadratic extrapolation to $a \rightarrow 0$
yields $B_c=0.1659$. A much more crucial issue  is the behaviour with
respect to $t$ and $y$. For that purpose we studied $\Epsilon_0 = \Epsilon_2$.
Indeed, this function shows a exponentially localized distribution around
the origin. After the appropriate scaling is done, we show in Fig.~1 the
lattice approximations to  $\log(\Epsilon_0 (t))$. Data for $N_s=4,6$ are
plotted together and compared with a straight line $D-wt$. Apart from the
clear exponential fall off, one sees that the data for different $N_s$
compare nicely to within a few percent level.  At the origin, for example,
both data  differ by $4 \%$. This agreement is striking given the large
values of $a$ involved. The decay exponent $w$ was found to vary from
$6.50(6)$ to $6.65(5)$ for $N_s=4$ and $6$ respectively. Errors reflect
differences in fitting procedures, ranges, etc. 
We also looked to the joint $t$ and $y$ distribution. Our data suggests that
the action density  is indeed rotationally invariant in that plane, thus,
depending only on $r=\sqrt{t^2+y^2}$.

Hence, we have verified that we are actually dealing (to within the precision
of our data) with a self-dual continuum configuration, which is periodic 
with unit
period in the $x$ and $z$ directions and localized in the $t$ and $y$
directions. By considering more than one period in the $x,z$ directions
we  can obtain a new self-dual configuration with zero twist vectors (and hence,
obtainable by using periodic boundary conditions on the lattice) and a larger size.
The most important  feature being that its action density distribution is
concentrated around the $t=y=0$ plane. This is the world sheet of our (thick)
vortex. However, to  verify that we are indeed dealing with vortices, we
have to examine the behaviour of the Wilson loop around this object.
The relevant loops, with our choice of
coordinates, are the $0-2$ loops. Hence, we evaluated  square loops of increasing
linear size $r$ centered at $t=y=0$: ${\cal L}(r,x,z)=\frac{1}{2} Tr( W_{0
2}(r \times r))$. The results are  given in Fig.~2. We show the two extreme
cases $x=z=0$ and $x=z=\frac{1}{2}$; the rest of values give curves lying in
between.
Again data for $N_s=4$ and $N_s=6$ are plotted
together to give an idea of the size of O(a) effects.
One sees that ${\cal L}(r)$ becomes 1 for small r, but tends to -1
as r grows, as expected. It is, of course, this property that makes these
configurations effective in disordering the Wilson loop and approaching
Confinement, just as  thin $Z_2$ fluxons in $Z_2$ gauge theories.


\section{Considerations on Confinement}
 In the previous section, we have shown that there exist self-dual
 configurations corresponding to a vortex wall aligned along one
  plane. Taking this plane to be the $z-t$ plane (it could be any), we
  can interpret this wall as the world sheet of an infinitely long  vortex
 line located around the z axis. This is similar to the case of the abelian
 vortex solution found by Nielsen and Olesen~\cite{abvortex}. Just as for
that case, one can adiabatically deform the string to form a finite closed
loop in 3-space. Similarly, the corresponding infinite cylinder in space-time
can be changed into a 2-dimensional torus representing a vortex-antivortex
loop, or (possibly) into a sphere representing the creation, propagation
and annihilation of a vortex ring. However, our non-abelian vortex solution
is not static as its  abelian partner, but {\em breathes} at a fixed pace.
The time-period, whose magnitude  determines the thickness  of the wall, 
can be any positive real number.
Similarly, it is not constant but periodic in the z direction (It is a
breathing caterpillar). The fact that
the periods in $z$ and $t$ are equal, as in our numerical example, is an
unnecessary  constraint. There are also solutions with different periods in 
both directions, as we explicitly verified numerically. We simply focused
on the most symmetric case.

Much less obvious is
the fact that it is also possible to depart from the periodic arrangement.
To see this, just  put back our solution in a torus with  a size which is
a multiple of the periods. Then, the Atiyah-Singer index theorem tell us that
there are exactly $4N$ deformations of the solution that lead back to new
self-dual solutions with equal action. The integer $N$ is the number of
unit cells (equal to the number of $Q=\frac{1}{2}$ lumps,  or to the total
action divided by $4 \pi^2$). This is just the right number of degrees of
freedom required to  move freely the lump centers away from the periodic
arrangement. So it is quite plausible,
that for small deformations a configuration looks
 like a picture of a 2-dimensional solid at finite temperature, with the
ions displaced from the periodic array of equilibrium positions. 
One does not know, however, 
what could happen for large deformations. 

In conclusion, the vortex solution that we have found has some features
which are non-generic, artifacts of the method used to construct it. This is typical
of the search of classical solutions, where normally the most symmetric
ones are more easily found or constructed. We do not know if, in the same
spirit, there exist  solutions which cannot be constructed in terms
of $Q=\frac{1}{2}$ lumps. For example, the new caloron solutions mentioned
earlier are made of 2 lumps with $Q=\tau$ and $1-\tau$ for any $0 \le \tau
\le \frac{1}{2}$. This is an interesting point which is  harder to investigate
with our method.

Finally, we would like  to comment upon the possible relevance of these solutions
for the QCD vacuum and Confinement. A good deal of strategies and
ideas have been put forward to understand Confinement. However, it is
important to realize that all the ideas are not necessarily conflicting. In
 the most popular approach~\cite{super}, one maps the non-abelian problem into an
abelian model (with monopoles) by means of the abelian projection,  and uses the 
well-understood  abelian dual-superconductor confinement mechanism to
describe  the non-abelian  one. This is indeed very appealing and probably 
correct, however, although not frequently realized, this is not the whole
story. Having abelian monopoles does not imply having Confinement, as the 
4-dimensional compact abelian case shows. There is something special about 
the non abelian theory which makes the condensation work all the way to zero 
bare coupling. It is precisely the mechanism that drives the condensation
that some of us are trying to understand.  As an 
analogy, one can take the case of Superconductivity. Realizing that the
Bose-Einstein condensation of charged particles underlies the phenomenon,
does not eliminate the necessity to search for  the   microscopic mechanism 
that drives the condensation:  BCS  theory.

Our group has advocated that self-dual structures, like the configurations
presented here,  are indeed responsible for Confinement. Self-dual
configurations are  called multi-instanton in the literature,  
but this has not to be confused with an array of independent instantons. 
One may say, that when instantons are pushed together 
tightly, they {\em merge} into new structures with special properties (see
the comments on Ref.~\cite{polyakov}).
The vortex configuration
presented here shows an example of this. For example, the topological charge
per lump is $\frac{1}{2}$ rather than 1. This may or may not be a generic
feature, but it is very appealing to realize that the number of degrees
of freedom of the moduli space equals the number of degrees of freedom of a
liquid of $Q=\frac{1}{2}$  lumps. Energy-entropy arguments of why these
structures would become condensed, were given in our  paper
Ref.~\cite{Investigating}. A crucial point raised by the advocators of the 
vortex mechanism~\cite{lat98}, is what is the  space-time
dimensionality of the structures responsible for Confinement. This  
is an important question that the lattice community has to settle. However, there
are self-dual structures of all dimensionality: from calorons(also called 
BPS monopoles or dyons) or their constituent monopoles, to the isotropic model 
of Ref.~\cite{Investigating}, passing through the vortex solutions of this paper. 
To complicate the picture,
ordinary isolated instantons are also expected to be present, although their 
role in Confinement  is presumably limited.

\section*{Acknowledgements}
The present work was financed by CICYT under grant AEN97-1678.

\newpage

\begin{figure}

\caption{We plot the logarithm of the the energy profile $\Epsilon_{0}$(t) for our
solution. Open circles and full squares correspond to lattice spacing
values of  $a=0.1666$ and $0.25$ respectively. The straight line is given by
$5.42 - 6.5 \, t$.}

\vbox{ \hskip -1cm \hbox{  \epsfxsize=450pt \hbox{\epsffile{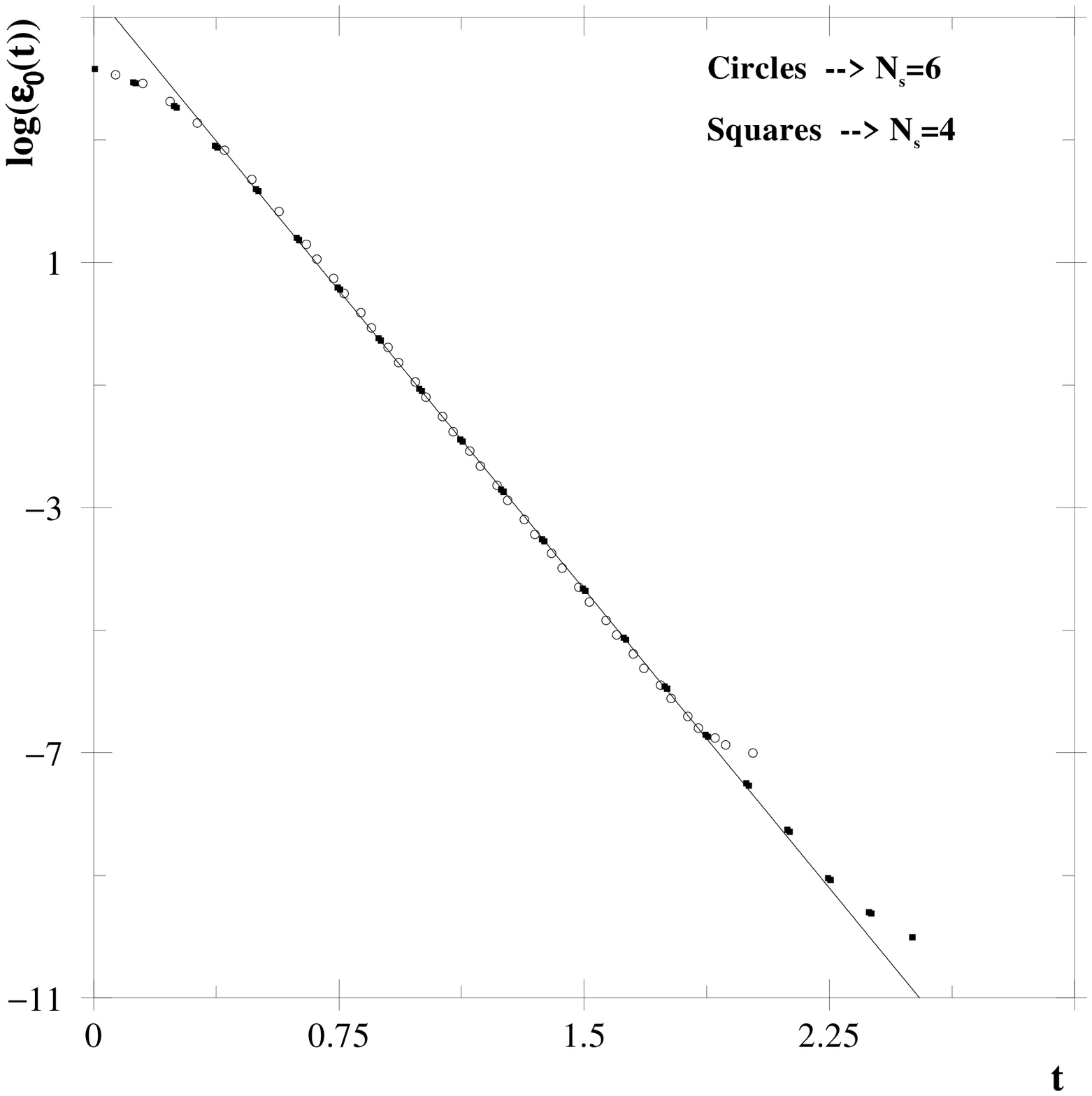} }
}  
     }
     
     \label{profile }
     
     \end{figure}

     \newpage
     
     \begin{figure}

     \caption{ We plot the value of the trace of an $r \times r$  Wilson loop
     centered around the vortex ${\cal L}(r,x,z)$. The different points are
 labeled in the figure, with  {\em max} corresponding to $x=z=0$ and  {\em
min} to  $x=z=\frac{1}{2}$.}
     
     \vbox{ \hskip -1cm \hbox{  \epsfxsize=450pt
\hbox{\epsffile{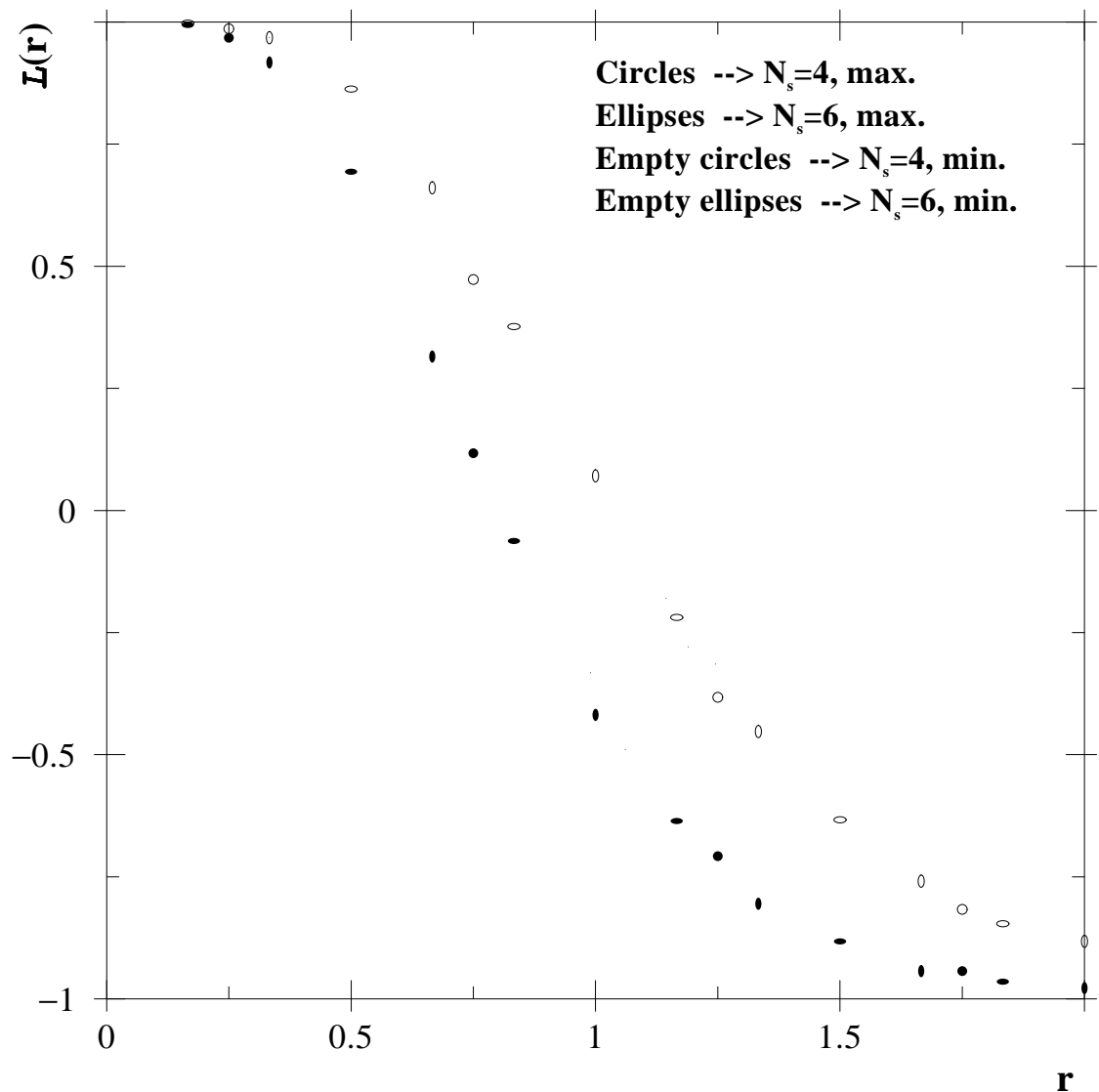} }
}  
     }
     
          \label{wilsonloop}
	  
	       \end{figure}

	            \newpage

\end{document}